\documentclass[sigconf,authorversion=true,nonacm=true]{acmart}
\AtBeginDocument{%
  }

\usepackage{tabularx}
\usepackage{booktabs}   

\newcommand{\gender}{\textit{Gender}}
\newcommand{\age}{\textit{Age}}
\newcommand{\race}{\textit{Race}}
\newcommand{\education}{\textit{Education}}
\newcommand{\occupation}{\textit{Occupation}}
\newcommand{\lang}{\textit{Language}}
\newcommand{\health}{\textit{Health}}
\newcommand{\economic}{\textit{Economic}}
\newcommand{\region}{\textit{Region}}
\newcommand{\sick}{\textit{\#Sick}}
\newcommand{\vaccinated}{\textit{Vaccinated}}
\newcommand{\source}{\textit{Source}}
\newcommand{\trust}{\textit{Trust}}
\newcommand{\fear}{\textit{Fear}}
\newcommand{\wave}{\textit{Wave}}
\newcommand{\trend}{\textit{Trend}}

\newif\ifannote
\ifannote
    \newcommand{\anninsert}[2]{{\color{#1}#2}}
    \newcommand{\anncomment}[3]{{\color{#1}[#2: #3]}}
\else
    \newcommand{\anninsert}[2]{#2}
    \newcommand{\anncomment}[3]{}
\fi

\newcommand{\af}[1]{\anninsert{magenta}{#1}}

\begin{document}

\title{What Causes COVID-19 Fear? General Drivers of Fear During a Health Crisis}


\author{D. Baccega$^{1}$, P. Castagno$^{1}$, A. Fernández Anta$^{2,3}$, J. M. Ramirez$^{3}$, M. Sereno$^{1}$}

\affiliation{%
  \institution{$^{1}$University of Turin}
  \city{Turin}
  \country{Italy}
}

\affiliation{%
  \institution{$^{2}$IMDEA Software Institute}
  \city{Madrid}
  \country{Spain}
}

\affiliation{%
  \institution{$^{3}$IMDEA Networks Institute}
  \city{Madrid}
  \country{Spain}
}

%\author{Daniele Baccega}
%\email{daniele.baccega@unito.it}
%\orcid{0009-0003-7331-8634}
%\author{Paolo Castagno}
%\email{paolo.castagno@unito.it}
%\orcid{0000-0002-1349-1844}
%\affiliation{%
%  \institution{University of Turin}
%  \city{Turin}
%  \country{Italy}
%}
%
%\author{Antonio Fernández Anta}
%\email{antonio.fernandez@imdea.org}
%\orcid{0000-0001-6501-2377}
%\author{Juan Marcos Ramirez}
%\email{juan.ramirez@imdea.org}
%\orcid{0000-0003-0000-1073}
%\affiliation{%
%  \institution{IMDEA Networks Institute}
%  \city{Madrid}
%  \country{Spain}
%}

%\author{Matteo Sereno}
%\email{matteo.sereno@unito.it}
%\orcid{0000-0002-5339-3456}
%\affiliation{%
% \institution{University of Turin}
%  \city{Turin}
%  \country{Italy}
%}

\renewcommand{\shortauthors}{Baccega et al.}

\begin{abstract}
The COVID-19 pandemic triggered not only a global health crisis but also an \textit{infodemic}, where exposure to heterogeneous information sources influenced public emotional responses. In this work, we investigate the determinants of self-reported fear of infection using data from the Delphi US CTIS survey. In particular, we analyze how demographic variables, epidemiological conditions, and exposure to different information sources shape fear levels.
We introduce a Probabilistic Causal Model to estimate causal relationship strengths, identifying the variables that most strongly influence fear.
Our results indicate that exposure to information sources accounts for a greater proportion of variance in fear than demographic and epidemiological variables do. We further compute the Average Treatment Effect to quantify the impact of different information sources on fear. After causal adjustment, institutional and expert-driven sources are associated with increased fear levels, whereas politicians, religious leaders, and alternative information channels are associated with reduced fear. These findings highlight both the central role of the information ecosystem in shaping emotional responses during public health crises and the value of causal inference approaches for studying behavioral responses to pandemics.
\end{abstract}



\keywords{COVID-19, Fear, Causal inference, Sources of information, Survey-based analysis}


\maketitle

%\vspace{-0.5cm}
\section{Introduction}\label{sec:intro}
The COVID-19 pandemic represented a global health crisis, while simultaneously it generated an \textit{infodemic}:
a rapid surge of information from heterogeneous sources~\cite{cinelli2020covid}.
Individuals relied on news media, social platforms, government announcements, and public health organizations for guidance.
The credibility, consistency, and framing of these sources shaped both public understanding of the threat and emotional responses across populations~\cite{jones2021understanding,lu2021source}.

In this context, fear emerged as one of the main psychological responses to the pandemic~\cite{coelho2020nature}.
This emotion was driven not only by the objective health threat, but also by the volume, tone, and perceived trustworthiness of consumed information.
The impact of information sources varied across demographic groups: age, education, gender, and socioeconomic status shaped how individuals selected, interpreted, and trusted information~\cite{nino2021race}.
These disparities highlight the need to account for both information characteristics and recipient profiles when evaluating the psychological effects of information exposure.\\
%\vspace{-0.01truecm}
\textbf{Contributions.}
This work provides three main contributions. First, we introduce a causal inference framework combining epidemiological indicators, demographic information, and information-source exposure to study fear dynamics during the COVID-19 pandemic. Second, we investigate how these demographic, epidemiological, and informational factors contribute to fear, emphasizing the need to distinguish causal effects from statistical associations. Third, by fitting a Probabilistic Causal Model (PCM) and using variance decomposition of arrow strengths and Average Treatment Effect (ATE) estimation, we quantify the role of information sources in shaping fear, showing how communication channels can amplify or mitigate emotional responses during public health crises.
%This work provides three main contributions. First, we introduce a causal inference framework combining epidemiological indicators, demographic information, and information-source exposure to study fear dynamics during the COVID-19 pandemic. Second, we investigate how different demographic, epidemiological, and informational factors contribute to fear, highlighting the importance of distinguishing causal effects from simple statistical associations. Third, through variance decomposition and Average Treatment Effect (ATE) estimation, we quantify the role of information sources in shaping fear, showing how different communication channels can either amplify or mitigate emotional responses during large-scale public health crises.

\begin{table*}[!ht]
  \centering
  \tiny
  \caption{Selected questions and responses from the Delphi US CTIS survey, shown with original survey labels and the corresponding variable names used throughout the paper. Only the responses included in the analysis are presented, aggregated as indicated. All variables are binary or categorical. % For instance, for gender, non‑binary and self‑described responses were excluded due to a limited number of responses.
   }
  \vspace{-0.4cm}
  \begin{tabularx}{\textwidth}{p{6cm}|X|p{0.7cm}|c}
    \toprule
    \textbf{Question} & \textbf{Responses} & \textbf{Name} & \textbf{Measured}\\
    \midrule
    \midrule
    \textit{(D1)} What is your gender? & \textit{\textbf{(1)}} Male, \textit{\textbf{(2)}} Female & \gender{} & {\color{green}$\checkmark$}\\ 
    \textit{(D2)} What is your age? & \textit{\textbf{(1)}} 18-34 years, \textit{\textbf{(3)}} 35-64 years, \textit{\textbf{(6)}} 65+ years & \age{} & {\color{green}$\checkmark$} \\
    \textit{(D7)} What is your race? Please select all that apply. & \textit{\textbf{(1)}} American Indian or Alaska Native, \textit{\textbf{(2)}} Asian, \textit{\textbf{(3)}} Black or African American, \textit{\textbf{(4)}} Native Hawaiian or other Pacific Islander, \textit{\textbf{(5)}} White, \textit{\textbf{(6)}} Some other race & \race{} & {\color{red}$\times$}\\
    \textit{(D8)} What is the highest degree or level of school you have completed? & \textit{\textbf{(1)}} High school or less, \textit{\textbf{(2)}} Bachelor / Master degree, \textit{\textbf{(3)}} Postgraduate & \education{} & {\color{green}$\checkmark$}\\
    \textit{(D9)} In the past 4 weeks, did you do any kind of work for pay? & \textit{\textbf{(1)}} Yes, \textit{\textbf{(2)}} No & \occupation{} & {\color{green}$\checkmark$}\\
    \textit{(D12)} What language do you speak most often at home? & \textit{\textbf{(1)}} English, \textit{\textbf{(2)}} Other & \lang{} & {\color{green}$\checkmark$} \\
    \textit{(C1)} Have you ever been told by a doctor, nurse, or other health professional that you have any of the following medical conditions? Please select all that apply. & \textit{\textbf{(1)}} Cancer (other than skin cancer), \textit{\textbf{(2)}} Heart attack, heart disease, or other heart condition, \textit{\textbf{(3)}} High blood pressure, \textit{\textbf{(4)}} Asthma, \textit{\textbf{(5)}} Chronic lung disease such as COPD, chronic bronchitis, or emphysema, \textit{\textbf{(6)}} Kidney disease, \textit{\textbf{(7)}} Type 1 diabetes, \textit{\textbf{(8)}} Type 2 diabetes, \textit{\textbf{(9)}} Weakened or compromised immune system, \textit{\textbf{(10)}} Obesity, \textit{\textbf{(11)}} None of these & \health{} & {\color{red}$\times$}\\
    \textit{(C15)} How worried are you about your household's finances for the next month? & \textit{\textbf{(1)}} Not worried at all, \textit{\textbf{(2)}} Not too worried, \textit{\textbf{(3)}} Somewhat worried, \textit{\textbf{(4)}} Very worried & \economic{} & {\color{red}$\times$}\\
    \textit{(A3)} What is your current ZIP code? & Open answer (we map ZIP codes with U.S. states and U.S. states into regions, i.e. Northeast, Midwest, South, West) & \region{} & {\color{green}$\checkmark$}\\
    \textit{(A4)} How many additional people in your local community do you personally know who are sick with a fever, along with at least one other symptom from the above list? & Open answer (we keep only values less than or equal to 100, a threshold which is conservative and filters out unrealistic responses, like those reporting millions) & \sick{} & {\color{green}$\checkmark$}\\
    \textit{(V1)} Have you had a COVID-19 vaccination? & \textit{\textbf{(1)}} Yes, \textit{\textbf{(2)}} No & \vaccinated{} & {\color{green}$\checkmark$} \\
    \textit{(I5)} In the past 7 days, from which of the following sources have you received news and information about COVID-19? &
          \textit{\textbf{(1)}} Doctors and other health professionals you go to for medical care, 
          \textit{\textbf{(2)}} Scientists and other health experts, 
          \textit{\textbf{(3)}} Centers for Disease Control (CDC), 
          \textit{\textbf{(4)}} Government health authorities or officials, 
          \textit{\textbf{(5)}} Politicians, 
          \textit{\textbf{(6)}} Journalists, 
          \textit{\textbf{(7)}} Family and friends,
          \textit{\textbf{(8)}} Religious leaders, 
          \textit{\textbf{(9)}} None of the above (other sources not listed here) & \source{} & {\color{green}$\checkmark$}\\
    \textit{(I6)} How much do you trust the following sources to provide accurate news and information about COVID-19? &                          
          For each \textit{I5}: \textit{\textbf{(1)}} Do not trust, \textit{\textbf{(2)}} Somewhat trust, \textit{\textbf{(3)}} Trust & \trust{} & {\color{green}$\checkmark$}\\
    \textit{(G1)} How much do you worry about catching COVID-19? &
          \textit{\textbf{(1)}} Not at all, 
          \textit{\textbf{(2)}} A little,
          \textit{\textbf{(3)}} A moderate amount,
          \textit{\textbf{(4)}} A great deal & \fear{} & {\color{green}$\checkmark$}\\
    \bottomrule
  \end{tabularx}
  \label{tab:selected_questions}
  \vspace{-0.4cm}
\end{table*}

\section{Related work}\label{sec:relatedwork}
Fear emerged as a central psychological response during the COVID-19 pandemic, shaped more by subjective perceptions and social transmission than by epidemiological indicators. 
Cross-national longitudinal research found that roughly 23\% of fear variance was explained by micro-level psychological factors (including worries about shortages, perceived vulnerability, germ aversion, and infections in close social circles), while macro-level indicators such as case counts, mortality, and restrictions contributed minimally~\cite{eder2021predicting}.

Social media analyses showed that fear peaked during early outbreak phases, and that %differed substantially by gender, with women expressing fear more frequently than men~\cite{chai2022measuring}. In fact, 
females, healthcare workers, lower-educated, and married individuals showed higher fear levels~\cite{chai2022measuring,oliveira2023beyond,mertens2022fear,doshi2021assessing}.
\af{This had impact in the evolution of the pandemic.} Agent-based models showed that declining fear can trigger new disease waves by reducing protective behavior~\cite{retzlaff2022fear}. On the other end, early population-level fear modestly predicted later vaccination willingness but was insufficient to explain vaccine hesitancy.

Causal inference studies revealed bidirectional relationships between media exposure, interpersonal communication, and fear. Media consumption and perceived danger increased fear, while elevated fear motivated additional information seeking~\cite{li2022fear}.
Despite extensive research, the influence of different information sources on fear across demographic groups remains poorly characterized. Our study addresses this gap through causal inference methods evaluating how demographics, information sources, vaccination, and other variables shape fear responses.
\section{Data}\label{sec:data}
We used two COVID-19 data sources: official data from the COVID-19 Data Hub~\cite{covid-19-1, covid-19-2} and survey data from the Delphi US CTIS~\cite{delphi_covidcast_2020}. From the Data Hub, we extracted daily fatalities and confirmed cases to estimate infections via Sybil~\cite{baccega2024enhancing}, reconstructing a Susceptible-Infected-Recovered-Deceased-Susceptible (SIRDS) model over the study period assuming a 14-day average recovery time.

The Delphi US CTIS survey collected approximately 50,000 daily responses from Facebook users in the U.S. and was administered by Carnegie Mellon University under Facebook’s Data for Good program~\cite{facebook,astley2021global,fan2020university}. It includes self-reported information on symptoms, testing, vaccination, behaviors, trust in information sources, and fear. These data were used to estimate prevalence trends across U.S. states while accounting for direct and indirect reporting~\cite{Srivastava2024Nowcasting}. Access was obtained through an agreement with CMU, UMD, and Facebook.
Table~\ref{tab:selected_questions} reports the subset of Delphi US CTIS questions considered in this study (the full questionnaire is available at~\cite{delphi_ctis}) Data for \race{}, \health{}, and \economic{} variables are not available due to dataset limitations and inconsistencies (see Section~\ref{sec:limitations} for details).

Our analysis covers May 2021 to June 2022 across all U.S. states, combining epidemiological indicators with self-reported behavioral and psychological information. Although these sources are biased toward individuals with stable internet access, particularly active Facebook users, such biases are less pronounced in U.S. states with widespread connectivity, still providing valuable insights into large-scale behavioral trends during the COVID-19 pandemic.

\section{Model and Methods}\label{sec:methods}
\begin{figure}[!ht]
    \centering
    \vspace{-0.4cm}
    \caption{Causal inference model used.}
    \includegraphics[width=1\linewidth]{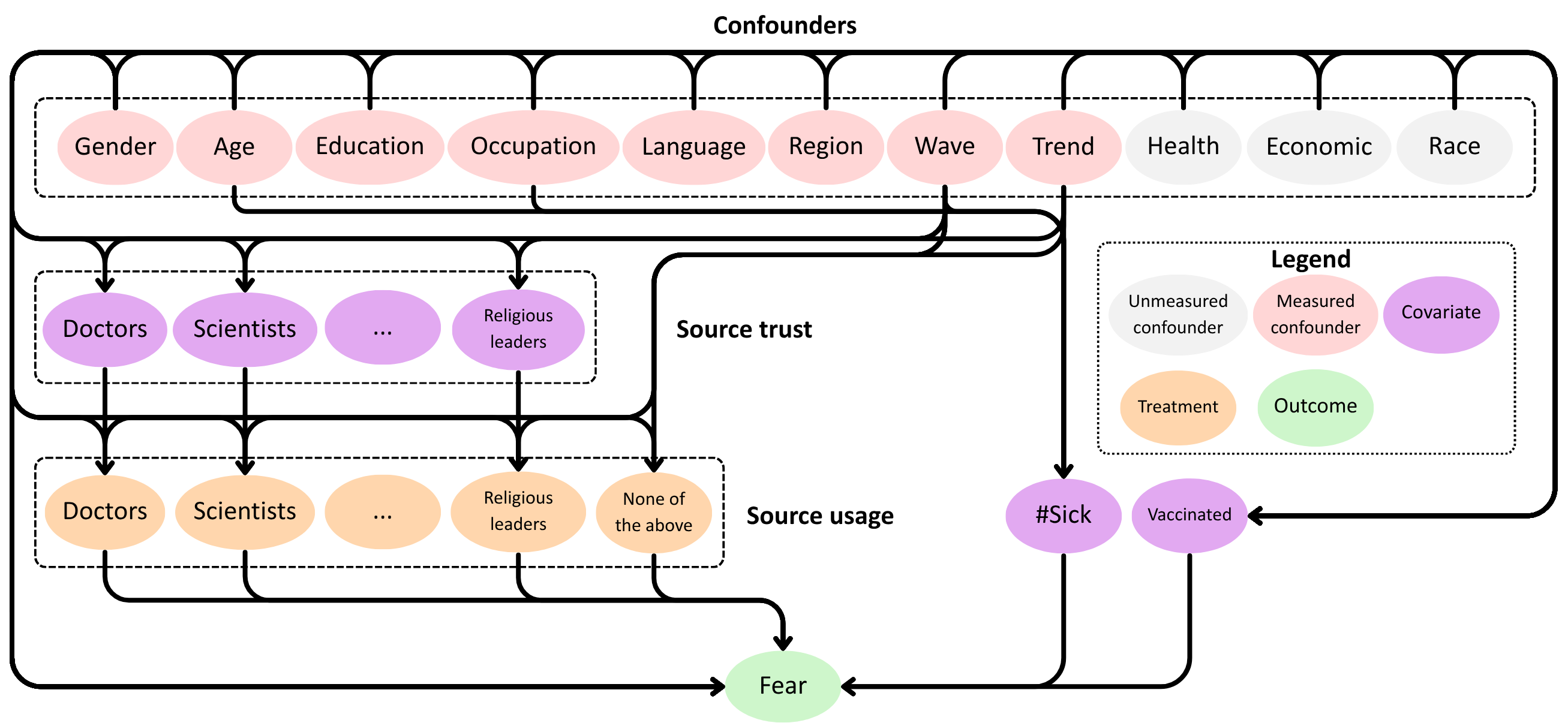} 
    \label{fig:cm}
    \vspace{-0.8cm}
\end{figure}

To define the causal inference model shown in Figure~\ref{fig:cm}, we use the DoWhy package~\cite{dowhy,JMLR:v25:22-1258}, a Python library integrating graphical causal models with statistical methods to specify assumptions, identify estimands, estimate causal effects, and perform robustness checks through refutation tests.
The Directed Acyclic Graph (DAG) encodes the assumed causal relationships among demographic and contextual confounders, trust in information sources, source usage, epidemiological variables, and fear outcomes.

In particular, the model represents how demographic characteristics (\gender, \age, \education, \occupation, \lang, and \region), together with pandemic-specific variables such as the COVID-19 \wave~ (in the period considered, we have two waves:Delta and Omicron) and epidemiological \trend~ (which maps the derivative of the number of infected individuals: -1 for values $\leq 0$ and 1 for values $> 0$), jointly influence usage of information sources and fear levels.

The treatment variables consist of nine binary \source~ indicators obtained through one-hot encoding of the original source-usage variable, while the outcome variable is COVID-19 \fear.
The DAG also includes other covariates: trust in information sources (\trust{}), and two epidemiological ones which directly contribute to \fear{}: the number of infected individuals known locally (\sick{}) and vaccination status (\vaccinated).
We additionally acknowledge potential unmeasured confounders such as health status (\health{}), economic conditions (\economic{}), and race (\race{})---see Section~\ref{sec:limitations} for details.

Since demographic and contextual variables affect both \source{} and \fear{}, naive associations between source usage and fear may be confounded.
The DAG, therefore, defines the adjustment set required to identify the causal effect of information-source usage on fear.
Rather than modeling calendar time explicitly, temporal dynamics are captured through the categorical COVID-19 \wave{} indicator and the epidemiological \trend{} variable, which reflects whether infection rates are increasing or decreasing. This formulation distinguishes periods of pandemic expansion and contraction.

DoWhy was primarily used to estimate the causal relationships represented in the DAG by fitting a PCM to the observed data.
The PCM enabled estimation of arrow strengths while preserving the conditional dependencies specified by the graph.
We additionally computed the ATE for each source variable to quantify the average causal effect of using a specific source on reported fear levels, after adjustment for DAG-identified confounders.
Whereas we use variance to quantify causal arrow strength (i.e., how much a parent node influences its child node), the ATE measures, under the assumptions encoded in the graph, the expected change in \fear{} associated with exposure to a given information source relative to non-exposure.
% Finally, DoWhy’s refutation procedures were used to evaluate the robustness of the estimated causal effects under alternative modeling assumptions and potential hidden confounding.

\section{Results}
%========================================
% Cramer's V
%========================================
\begin{table}[t]
\centering
\caption{Top-20 pairwise associations with Cramér's $V$.}
\vspace{-0.4cm}
\label{tab:cramers_v}
\tiny
\begin{tabular}{p{3.2cm} p{3.2cm} c}
\toprule
\textbf{Variable 1} & \textbf{Variable 2} & \textbf{Cramér's $V$} \\
\midrule
\fear{} & \vaccinated{} & 0.345 \\
Scientists / health experts \source{} usage & Scientists / health experts \source{} trust & 0.320 \\
Journalists \source{} usage & Journalists \source{} trust & 0.311 \\
CDC \source{} usage & CDC \source{} trust & 0.303 \\
Government authorities \source{} usage & Government authorities \source{} trust & 0.260 \\
\fear{} & Other / none of the above \source{} usage & 0.242 \\
Religious leaders \source{} usage & Religious leaders \source{} trust & 0.207 \\
\fear{} & CDC \source{} usage & 0.204 \\
\education{} & Scientists / health experts \source{} usage & 0.187 \\
\fear{} & \gender{} & 0.178 \\
\education{} & CDC \source{} usage & 0.169 \\
\fear{} & Scientists / health experts \source{} usage & 0.167 \\
Family and friends \source{} usage & Family and friends \source{} trust & 0.161 \\
\fear{} & \lang{} & 0.159 \\
\gender{} & CDC \source{} trust & 0.153 \\
\fear{} & Government authorities \source{} usage & 0.152 \\
\age{} & \vaccinated{} & 0.149 \\
Doctors \source{} usage & Doctors \source{} trust & 0.143 \\
\education{} & Other / none of the above \source{} & 0.133 \\
\education{} & Government authorities \source{} usage & 0.131 \\
\bottomrule
\end{tabular}
\vspace{-0.4cm}
\end{table}

% \begin{table}[t]
% \centering
% \caption{Pairwise associations with Cramér's $V$.}
% \vspace{-0.4cm}
% \label{tab:cramers_v}
% \tiny
% \begin{tabular}{p{3.2cm} p{3.2cm} c}
% \toprule
% \textbf{Variable 1} & \textbf{Variable 2} & \textbf{Cramér's $V$} \\
% \midrule
% \fear{} & \vaccinated{} & 0.345 \\
% Scientists / health experts \source{} usage & Scientists / health experts \source{} trust & 0.320 \\
% Journalists \source{} usage & Journalists \source{} trust & 0.311 \\
% CDC \source{} usage & CDC \source{} trust & 0.303 \\
% Government authorities \source{} usage & Government authorities \source{} trust & 0.260 \\
% \fear{} & Other / none of the above \source{} & 0.242 \\
% Religious leaders \source{} usage & Religious leaders \source{} trust & 0.207 \\
% \fear{} & CDC \source{} usage & 0.204 \\
% \fear{} & Scientists / health experts \source{} usage & 0.167 \\
% Family and friends \source{} usage & Family and friends \source{} trust & 0.161 \\
% \fear{} & Government authorities \source{} usage & 0.152 \\
% Doctors \source{} usage & Doctors \source{} trust & 0.143 \\
% \fear{} & Doctors \source{} usage & 0.107 \\
% \fear{} & Journalists \source{} usage & 0.102 \\
% \fear{} & \sick{} & 0.083 \\
% \fear{} & Family and friends \source{} usage & 0.071 \\
% \fear{} & Politicians \source{} usage & 0.061 \\
% Politicians \source{} usage & Politicians \source{} trust & 0.059 \\
% \fear{} & Religious leaders \source{} usage & 0.033 \\
% \bottomrule
% \end{tabular}
% \vspace{-0.4cm}
% \end{table}
%========================================
% Variance Explained
%========================================
\begin{table}[t]
\centering
\caption{Contributors to the variance explained in \fear{}.}
\vspace{-0.4cm}
\label{tab:arrow_strengths}
\tiny
\begin{tabular}{p{4cm} c c}
\toprule
\textbf{Parent Variable $\rightarrow$ Fear} & \textbf{Variance Explained} & \textbf{Variance Explained (\%)} \\
\midrule
\gender{} & 0.083 & 8.22\% \\
\vaccinated{} & 0.083 & 8.22\% \\
\sick{} & 0.065 & 6.44\%\\
Other / none of the above \source{} sources & 0.064 & 6.34\% \\
\occupation{} & 0.055 & 5.45\% \\
CDC \source{} usage & 0.049 & 4.85\% \\
Scientists / health experts \source{} usage & 0.041 & 4.06\%\\
Politicians \source{} usage & 0.030 & 2.97\%\\
Government authorities \source{} usage & 0.028 & 2.77\% \\
\education{} & 0.027 & 2.68\%\\
\region{} & 0.026 & 2.58\%\\
\age{} & 0.024 & 2.38\%\\
\lang{} & 0.021 & 2.08\%\\
Doctors \source{} usage & 0.019 & 1.88\%\\
Family and friends \source{} usage & 0.015 & 1.49\%\\
Journalists \source{} usage & 0.009 & 0.89\%\\
Religious leaders \source{} usage & 0.005 & 0.50\%\\
\midrule
\textbf{Total} & \textbf{0.644} & \textbf{63.80\%} \\
\bottomrule
\end{tabular}
\vspace{-0.4cm}
\end{table}
%========================================
% ATE
%========================================
\begin{table}[t]
\centering
\caption{ATE estimates of \source{} on \fear{}. 
%Positive values indicate increased fear, while negative values indicate reduced fear.
}
\label{tab:ate_results}
\vspace{-0.4cm}
\tiny
\begin{tabular}{p{7.3cm} c}
\toprule
\textbf{Treatment Variable} & \textbf{ATE} \\
\midrule
CDC & 0.227 \\
Scientists / health experts & 0.147 \\
Government authorities & 0.114 \\
Doctors / health professionals & 0.089 \\
% Number of sick people known locally & 0.027 \\
Family and friends & 0.027 \\
Journalists & -0.029 \\
Religious leaders & -0.141 \\
Politicians & -0.284 \\
Other / none of the above sources & -0.464 \\
% Vaccination status & -0.706 \\
\bottomrule
\end{tabular}
\vspace{-0.4cm}
\end{table}
To preliminarily assess relationships among the variables, we computed pairwise associations using Cramér’s V. Table~\ref{tab:cramers_v} reveals weak to moderate associations between several variable pairs. Notably, \trust{} and the corresponding \source{} variables show consistent positive associations, indicating that individuals who trust a given source also tend to rely on it.
% To preliminarily assess relationships among the considered variables, we computed pairwise associations using Cramér’s V. Table~\ref{tab:cramers_v} shows moderate/low correlations between several variables, particularly between trust in information sources and the corresponding source-usage variables, suggesting that individuals who trust a source also tend to rely on it.
%
\fear{} exhibits moderate associations with \vaccinated{} and, to a lesser extent, with \gender{}, \lang{}, and \source{} from the \emph{CDC}, \emph{scientists and health experts}, \emph{government authorities}, the \emph{journalists}, and \emph{other unspecified sources}. However, these associations are not causal because demographic and epidemiological variables may jointly affect both information-source usage and fear. This motivates the need for causal inference methods.

To address these limitations, we estimated the PCM described in Section~\ref{sec:methods}. Table~\ref{tab:arrow_strengths} reports the variance explained by the main causal arrows targeting \fear{}. Overall, the variables included in the model account for about 64\% of the variance in fear (which is equals to 1.0093), indicating that the considered demographic, epidemiological, and informational variables collectively capture a substantial portion of fear variability.

Information-related variables emerge as the dominant explanatory component. Combined, information-source variables explain more than 25\% of the total variance in \fear{}, exceeding the contribution of any other variable group in the model. In particular, receiving information from \emph{other unspecified sources}, the \emph{CDC}, \emph{scientists and health experts}, \emph{politicians}, and \emph{government authorities} explains a substantial portion of fear variance. \af{Among individual variables, \gender{} and \vaccinated{} contribute the most} (a result supported by~\cite{chai2022measuring} and~\cite{mertens2022fear}). \sick{} also contributes strongly, confirming that epidemiological conditions shape fear levels. Other demographic variables, such as \occupation{}, \education{}, \age,{} and \lang{}, further explain part of the variance, suggesting heterogeneous psychological responses across population groups.

% Overall, fear appears to be shaped by epidemiological conditions, demographic characteristics, and information exposure, with information-source variables playing a central role in explaining emotional responses during the pandemic.

Finally, we quantified the causal effect of each \source{} through ATE estimation using propensity score stratification. Table~\ref{tab:ate_results} reveals substantial heterogeneity across sources. Institutional and expert-driven sources, including the \emph{CDC}, \emph{scientists and health experts}, \emph{government authorities}, and \emph{doctors and health professionals}, are associated with increased fear levels. Conversely, relying on \emph{alternative or unspecified sources} (“None of the above”), \emph{politicians}, and \emph{religious leaders} is associated with reduced fear after adjustment for confounding.
Importantly, the ATE estimates differ substantially from the pairwise associations observed in the Cramér’s V analysis. Variables strongly associated with fear in the correlation analysis may exhibit weaker causal effects after adjustment for demographic and epidemiological confounding, while others emerge only after causal adjustment. This discrepancy highlights the importance of causal inference approaches when studying fear determinants during the COVID-19 pandemic.

\section{Limitations\label{sec:limitations}}
This work has several limitations related to data availability and survey structure. First, we could not include potentially important confounders such as race (\race{}), previous health conditions (\health{}), or detailed economic status (\economic{}). \race{} was
%excluded , 
\af{not available for ethical reasons},
while \health{} and \economic{} variables could not be reliably integrated because of dataset limitations and inconsistencies. Consequently, unmeasured confounding may still bias part of the estimated causal effects despite DAG-based adjustment.

A second limitation is the absence of longitudinal user-level information. The Delphi US CTIS survey does not guarantee repeated observations of the same individuals and provides no persistent identifiers. Consequently, we could not construct a panel dataset or include lagged variables capturing delayed effects of information exposure or prior fear levels, preventing explicit modeling of temporal causal dependencies.
We explored aggregating observations into daily averages to introduce a temporal dimension. However, conditioning on the large set of demographic and contextual confounders produced subgroup sample sizes too small for reliable estimation. Directly modeling time was therefore not feasible within the adopted framework.

To partially address this limitation, we introduced two variables capturing epidemic evolution without explicit temporal indexing: the categorical \wave{} variable and the epidemiological \trend{} variable. The former distinguishes major pandemic phases, while the latter captures epidemic expansion or contraction through the derivative of estimated infections. Together, these variables provide a compact representation of epidemic dynamics despite the absence of longitudinal observations.
Within the DAG, \wave{} and \trend{} influence trust in information sources, information-source usage, vaccination behavior, and the perceived number of sick individuals, but not \fear{} directly. The assumption is that individuals do not directly observe epidemiological indicators; rather, information reaches them through media, institutions, social interactions, and other channels represented in the model. Although simplified, this assumption allows the model to isolate the indirect role of epidemic evolution through the information ecosystem.

\section{Conclusion}\label{sec:conclusion}
This study investigated the determinants of fear of COVID-19 infection using epidemiological, behavioral, and demographic data from the Delphi US CTIS survey during the Delta and Omicron waves.

Using a Probabilistic Causal Model and the Average Treatment Effect estimation, we quantified the contribution of demographic, epidemiological, and informational factors to fear dynamics. Information sources emerged as the largest explanatory component of fear variance, exceeding the contribution of any other variable group. Institutional and expert-driven sources were associated with increased fear, whereas politicians, religious leaders, and alternative information channels were associated with reduced fear after causal adjustment.

Our findings suggest that fear was shaped not only by epidemiological conditions, but also by how the pandemic was communicated through the information ecosystem. The framing, credibility, and emphasis of information sources could amplify or mitigate emotional responses independently of underlying epidemiological conditions.

Methodologically, this work highlights the importance of causal inference approaches for studying behavioral responses during large-scale crises. By accounting for confounding variables and interdependencies among information channels, causal models provide a more robust framework for analyzing fear determinants than correlation analyses.

\section{Ethical declaration}
This work received ethical approval from the Ethics Board (IRB) of IMDEA Networks Institute on 2021/07/05. IMDEA Networks signed Data Use Agreements with Facebook and Carnegie Mellon University (CMU) 
%and the University of Maryland (UMD) 
to access data from 
%UMD project 1587016-3 (C-SPEC: Symptom Survey: COVID-19) and 
CMU project STUDY2020\_00000162 (ILI Community-Surveillance Study). The data were collected by CMU through the Global COVID-19 Trends and Impact Survey in partnership with Facebook. Informed consent was obtained from all participants. All methods were carried out in accordance with relevant ethical and privacy regulations.

\section{Availability of materials and data}
The COVID-19 Data Hub~\cite{covid-19-1, covid-19-2} is open-source and publicly available. The microdata of the Delphi US CTIS survey cannot be shared due to the Data Use Agreements signed with Facebook and Carnegie Mellon University (CMU).
\bibliographystyle{ACM-Reference-Format}
\bibliography{sample-base}

\end{document}